\documentclass[
	amsmath,
	amssymb,
	aps,
	prb,
	onecolumn,
]{revtex4-1}

\usepackage{graphicx}% Include figure files
\usepackage{dcolumn}% Align table columns on decimal point
\usepackage{bm}% bold math
\usepackage{amsmath}
\usepackage{amssymb}
\usepackage{color}
\usepackage{tikz}
\usepackage{pgfplots}
\newcommand{\surf}{\mathcal{S}}

\newcommand{\dS}{\textup{d}\surf}
\newcommand{\R}{\mathbb{R}}

\newcommand{\para}{\boldsymbol{X}}

\newcommand{\vnor}{\mathcal{V}}
\newcommand{\normal}{\boldsymbol{\nu}}

\newcommand{\shop}{\mathcal{B}}

\newcommand{\meanc}{\mathcal{H}}
\newcommand{\gaussc}{\mathcal{K}}

\newcommand{\areaPen}{\omega_{\text{a}}}
\newcommand{\area}{A}
\newcommand{\areaZero}{\area_0}

\newcommand{\inputTikzPic}[1]{\ifthenelse{\boolean{plotTikzPics}}{\input{#1}}{\fbox{\centering\begin{minipage}[t][5cm][t]{0.9\textwidth}\centering\textbf{\color{red}enable plotTikzPics}\end{minipage}}}}

\newcommand{\Grad}{\nabla}
\newcommand{\Div}{\operatorname{div}}%
\newcommand{\DivSurf}{\Div_{\!\surf}}%
\newcommand{\GradSurf}{\Grad_{\!\surf}}

\newcommand{\vecNabla}{\boldsymbol{\nabla}}

\newcommand{\levicivita}[2]{\vecNabla_{\!\!#1}{#2}}

\newcommand{\laplaceBeltrami}{\Delta_{\surf}}
\newcommand{\vecLaplace}{\boldsymbol{\Delta}}

\newcommand{\laplaceBochner}{\vecLaplace^{\!\textup{B}}}

\newcommand{\metric}{{\boldsymbol{g}}}

\newcommand{\deformationTensor}{\boldsymbol{d}}
\newcommand{\tangentVelocity}{\boldsymbol{v}}
\newcommand{\surfaceVelocity}{\boldsymbol{u}}

\newcommand{\surfaceInnerProduct}[2]{\left\langle #1 , #2 \right\rangle_\metric}

\newcommand{\pressure}{p}
\newcommand{\id}{\boldsymbol{\pi}}
\newcommand{\surfaceId}{\id_\surf}

\newcommand{\hTF}{\xi}
\newcommand{\TF}{\Omega_{\hTF}}
\newcommand{\veloTF}{\boldsymbol{V}}
\newcommand{\prsTF}{P}
\newcommand{\prsStressTF}{\boldsymbol{\Sigma}_{\prsTF}}
\newcommand{\paraTF}{\para_{\hTF}}

\renewcommand{\Re}{\textup{Re}}

\newcommand{\f}{\boldsymbol{f}}
\newcommand{\abbrevG}{g}

\newcommand{\Be}{\textup{Be}}

\newcommand{\intermediateVelocity}{\tangentVelocity^\star}

\newcommand{\tr}{\operatorname{tr}}

\newcommand{\insertColorbarHorizontal}[5]{
	\begin{minipage}{4.5cm}
		\begin{center}
			\begin{tikzpicture}
				\node (colorbar) at (0,0) {\includegraphics[width=3.6cm]{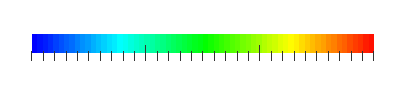}};
				\draw (0.0,0.35) node {\scriptsize #1};
				\draw (-1.53,-0.05) node[anchor=north] {\scriptsize #2};
				\draw (-0.53,-0.05) node[anchor=north] {\scriptsize #3};
				\draw (0.47,-0.05) node[anchor=north] {\scriptsize #4};
				\draw (1.47,-0.05) node[anchor=north] {\scriptsize #5};
			\end{tikzpicture}
		\end{center}
	\end{minipage}
}

\newcommand{\Ekin}{E}
\newcommand{\EkinT}{\Ekin_{\textup{T}}}
\newcommand{\EkinN}{\Ekin_{\textup{N}}}

\usepackage{ifthen}
\newboolean{forArxiv}
\setboolean{forArxiv}{true}

\ifthenelse{\boolean{forArxiv}}
{
	\newcommand{\eg}{e.~g.}
}
{
}

\begin{document}

%\newtheorem{lemma}{Lemma}
%\newtheorem{corollary}{Corollary}

% \ifthenelse{\boolean{forArxiv}}
% {
% }
% {
% }

\ifthenelse{\boolean{forArxiv}}
{
	
}
{
	\shorttitle{A numerical approach for fluid deformable surfaces} %for header on odd pages
	\shortauthor{S. Reuther et al.} %for header on even pages
}

\title{A numerical approach for fluid deformable surfaces}

\ifthenelse{\boolean{forArxiv}}
{
	\author
	{
		S. Reuther$^{1}$,
		I. Nitschke$^{1}$
		and 
		A. Voigt$^{1,2,3}$
	}
	\affiliation
	{
		$^{1}$
		Department of Mathematics, Technische Universit\"at Dresden, 01062 Dresden, Germany
		$^{2}$
		Center of Systems Biology Dresden (CSBD), Pfotenhauerstr. 108, 01307 Dresden, Germany
		$^{3}$
		Cluster of Excellence Physics of Life (PoL), Tatzberg 47/49, 01307 Dresden, Germany
	}
}
{
	\author
	{
	S. Reuther\aff{1},
	I. Nitschke\aff{1}
	\and 
	A. Voigt\aff{1,2,3}
		\corresp{\email{axel.voigt@tu-dresden.de}}
	}

	\affiliation
	{
	\aff{1}
	Department of Mathematics, Technische Universit\"at Dresden, 01062 Dresden, Germany
	\aff{2}
	Center of Systems Biology Dresden (CSBD), Pfotenhauerstr. 108, 01307 Dresden, Germany
	\aff{3}
	Cluster of Excellence Physics of Life (PoL), Tatzberg 47/49, 01307 Dresden, Germany
	}
}

\ifthenelse{\boolean{forArxiv}}
{
}
{
	\maketitle
}

\begin{abstract}
Fluid deformable surfaces show a solid-fluid duality which establishes a tight interplay between tangential flow and surface deformation. We derive the governing equations as a thin film limit and provide a general numerical approach for their solution. The simulation results demonstrate the rich dynamics resulting from this interplay, where in the presence  of  curvature  any  shape  change  is  accompanied  by  a  tangential  flow  and,  vice versa, the surface deforms due to tangential flow. However, they also show that the only possible stable stationary state in the considered setting is a sphere with zero velocity.  
\end{abstract}

\ifthenelse{\boolean{forArxiv}}
{
	\maketitle
}
{
}

\section{Introduction}

Fluid deformable surfaces are ubiquitous interfaces in biology, playing an essential role in processes from the subcellular to the tissue scale. Examples are lipid bilayers, the cellular cortex or epithelia monolayers. They all can be considered as fluidic thin sheets. From a mechanical point of view, they are soft materials exhibiting a solid-fluid duality: while they store elastic energy when stretched or bent, as solid shells, under in-plane shear they flow as viscous two-dimensional fluids. This duality has several consequences, it establishes a tight interplay between tangential flow and surface deformation. In the presence of curvature any shape change is accompanied by a tangential flow and, vice versa, the surface deforms due to tangential flow. The dynamics of this interplay strongly depends on the relation between fluid- and solid-like properties of the thin sheets. The growing interest on these phenomena in biology stays in contrast with available tools to numerically solve the governing equations. Even for surface fluids on stationary surfaces, where the governing equations are known since the pioneering work of \cite{Scriven_CES_1960}, numerical tools have only be developed recently, see \cite{Nitschkeetal_JFM_2012,Grossetal_JCP_2018} for simply-connected surfaces and \cite{Nitschkeetal_book_2017,Reutheretal_PF_2018,Fries_IJNMF_2018,Ledereretal_arXiv_2019} for general surfaces. The governing equations for fluid-deformable surfaces have been more recently derived in a different context, see \cite{Arroyoetal_PRE_2009, Salbreuxetal_PRE_2017,Miura_QAM_2018}, but never have been solved in a general setting. Recent approaches \cite{Torres-Sanchezetal_JFM_2019,Mietkeetal_PNAS_2019,Sahuetal_JCP_2020} are restricted to the Stokes limit, simply connected surfaces or axisymmetric settings. We overcome these limitations and provide a general numerical approach for fluid deformable surfaces.

The motivation to consider fluid deformable surfaces without the surrounding bulk phases results from the theoretical interest to explore them without any additional influence and the limit of a large Saffman-Delbr\"uck number. This number describes the relation between the viscosities of the surface and the typically less viscous bulk fluid and allows to decouple surface and bulk flow  \cite{Saffmanetal_PNAS_1975}.

The paper is structured as follows: In Section \ref{sec2} we sketch the derivation of the governing equations as a thin film limit and compare them with existing models for special cases. Section \ref{sec3} describes the numerical approach, which is based on evolution of geometric quantities and a generic finite element formulation for tensor-valued surface partial differential equations. Numerical examples demonstrating the tight interplay between tangential flow and surface deformation as well as convergence tests for the numerical approach are provided in Section \ref{sec4}. Conclusions are drawn in Section \ref{sec5}.

\section{Mathematical modeling}
\label{sec2}

We start from a slightly more general Navier-Stokes equation in the thin film $\TF(t) = \surf(t) \times [-\frac{\hTF}{2},\frac{\hTF}{2}] \subset \R^3$ with a regular evolving surface $\surf(t)$, film thickness $\hTF$ and surface normal $\normal$. It reads
\begin{align}
	\label{eq:ns1}
    \partial_t\veloTF + \levicivita{\veloTF}{\veloTF} &= \Div\prsStressTF + \frac{1}{\Re}\vecLaplace \veloTF \\
	\label{eq:ns2}
    \Div\veloTF &= 0
\end{align}
with $\veloTF$ velocity, $\Re$ Reynolds number, $\prsStressTF = -\prsTF\id - \phi\normal\otimes\normal$, where $\prsTF$ is the pressure, $\id = \mathbf{I} - \normal\otimes\normal$ and $\phi$ is an additional variable. The choice of $\phi = \prsTF$ results in the usual pressure gradient term $\Div\prsStressTF = -\Grad\prsTF$, which allows actions resulting from pressure differences in normal direction, whereas for $\phi = 0$ actions in normal direction resulting from pressure differences are omitted, leading to $\Div\prsStressTF = -\id\Grad\prsTF - \prsTF\meanc\normal$ with mean curvature $\meanc = \tr\shop$ and $\shop = -\GradSurf\normal$ the Weingarten mapping with covariant derivative $\GradSurf$. We consider a surface parametrization $\para$ and a thin film parametrization $\paraTF$ for which $\partial_t\para = \partial_t\paraTF|_{\surf} = \vnor\normal$, with $\vnor$ the normal velocity of the surface. This corresponds to an Eulerian description in the tangential space and a Lagrangian description in the normal direction. With slight modifications of the analysis in \cite{Nitschkeetal_PRF_2019} we obtain as the thin film limit $\hTF \to 0$ the surface Navier-Stokes equations for tangential and normal components of the surface velocity $\surfaceVelocity = \tangentVelocity + \vnor\normal$ and surface pressure $\pressure$
\begin{align}
    \label{eq:sns1}
   \!\!\!\! \surfaceId\partial_t\tangentVelocity + \! \levicivita{\tangentVelocity}{\tangentVelocity} \!-\! \vnor(\shop\tangentVelocity + \!\GradSurf\vnor) \!&=\! -\GradSurf\pressure \!+\! \frac{1}{\Re} \!\left( \laplaceBochner\tangentVelocity \!+\! \gaussc\tangentVelocity + \meanc\GradSurf\vnor \!-\! \vnor\GradSurf\meanc - 2\shop\GradSurf\vnor \right)\\
	\label{eq:sns2}
    \DivSurf\tangentVelocity \!&=\! \vnor\meanc \\
	\label{eq:sns3}
    \partial_t \vnor + 2\levicivita{\tangentVelocity}{\vnor} + \surfaceInnerProduct{\shop\tangentVelocity}{\tangentVelocity} \!&=\! - \partial_{\normal}\phi|_{\mathcal{S}} - \meanc( \pressure - \phi|_{\surf})  + \frac{2}{\Re} \!\left( \surfaceInnerProduct{\shop}{\GradSurf\tangentVelocity} - \vnor\|\shop\|^2 \right)
\end{align}
with $\levicivita{\tangentVelocity}{}$ the directional derivative, $\laplaceBochner$ the Bochner Laplacian, $\gaussc = \det\shop$ the Gaussian curvature and $\partial_{\normal}\phi|_{\surf}$ the normal derivative of $\phi$ in the thin film evaluated on the surface $\surf$. Eqs. \eqref{eq:sns1} and \eqref{eq:sns2} are independent of $\phi$. For given $\vnor$ these equations have also been previously derived by various approaches, see \cite{Arroyoetal_PRE_2009} (with corrected acceleration term \cite{Yavarietal_JNS_2016}) and \cite{Kobaetal_QAM_2017,Jankuhnetal_IFB_2018,Miura_QAM_2018,Nitschkeetal_PRF_2019}. For $\phi = 0$ also eq. \eqref{eq:sns3} is the same as the equation derived in \cite{Kobaetal_QAM_2017,Jankuhnetal_IFB_2018}. The surface Navier-Stokes equations \eqref{eq:sns1} - \eqref{eq:sns3} nicely show the tight coupling between $\tangentVelocity$ and $\vnor$ in the presence of curvature. Most prominently eq. \eqref{eq:sns2} forces any shape change to be accompanied by a tangent flow and the rate-of-deformation tensor $\deformationTensor = \frac{1}{2} \left( \GradSurf\tangentVelocity + (\GradSurf\tangentVelocity)^T \right) - \vnor\shop$, with
\begin{align*}
    2\DivSurf\deformationTensor &= \laplaceBochner\tangentVelocity + \gaussc\tangentVelocity + \GradSurf(\vnor\meanc) - 2\DivSurf(\vnor \shop) \\
    &= \laplaceBochner\tangentVelocity + \gaussc\tangentVelocity + \meanc\GradSurf\vnor - \vnor\GradSurf\meanc - 2\shop\GradSurf\vnor
\end{align*}
in eqs. \eqref{eq:sns1} and \eqref{eq:sns3} forces the surface to deform due to tangential flows. However, additional coupling terms are also present in the inertial terms.

Eqs. \eqref{eq:sns1} - \eqref{eq:sns3} assume fluid-like behavior in tangential and normal direction and can also be written in a more compact formulation for $\surfaceVelocity$. For $\phi = 0$ it reads
\begin{align}
    \label{eq:sns4}
    \partial_t\surfaceVelocity + \levicivita{\surfaceVelocity}{\surfaceVelocity} &= -\GradSurf\pressure + \frac{2}{\Re}\DivSurf\deformationTensor + \pressure\meanc\normal \\
	\label{eq:sns5}
    \DivSurf\surfaceVelocity &= 0
\end{align}
see  \cite{Jankuhnetal_IFB_2018}. However, this formulation hides the tight interplay between tangential and normal velocity components and is thus less suited to explore the resulting phenomena. Numerical approaches for eqs. \eqref{eq:sns1} - \eqref{eq:sns3} or \eqref{eq:sns4} and \eqref{eq:sns5} only exist for special cases. Most work, including also numerical analysis, is concerned with the Stokes limit on stationary surfaces $\vnor = 0$, see \eg\ \cite{Olshanskiietal_SIAMJSC_2018,Reusken_IMAJNA_2020}. For the surface Navier-Stokes equations in this situation see \eg\ \cite{Nitschkeetal_JFM_2012,Reutheretal_PF_2018,Fries_IJNMF_2018} and for their extension to evolving surfaces with prescribed $\vnor$ see \cite{Reutheretal_MMS_2015,Reutheretal_MMS_2018,Nitschkeetal_PRF_2019}. The Stokes limit of eqs. \eqref{eq:sns1} - \eqref{eq:sns3} or \eqref{eq:sns4} and \eqref{eq:sns5} corresponds to the classical model \cite{Scriven_CES_1960} and resamples, if coupled with bulk flow, with the Boussinesq-Scriven boundary condition in multiphase flow problems, see \eg\ \cite{Barrettetal_PRE_2015,Barrettetal_CMS_2015}.

We are only concerned with surface phenomena but are interested in an extended model, which in addition accounts for solid-like properties in normal direction. Such solid-fluid duality of fluid deformable surfaces is considered by supplementing the evolution equations with the contribution from a Helfrich energy $\frac{1}{\Be} \int_{\surf}(\meanc - \meanc_0)^2\dS$ to account for bending forces \cite{Helfrich_ZN_1973}, with $\Be$ the bending capillary number and $\meanc_0$ the spontaneous curvature. We will here only consider the case $\meanc_0 = 0$. 
%In addition we will enforce a volume constraint $\int_{\surf}\frac{\volPen}{2} (\frac{\vol - \volZero}{\vol})^2\dS$ with penalty parameter $\volPen > 0$, reference volume $\volZero$ and actual volume $\vol = \frac{1}{3} \int_{\surf} \langle \para, \normal \rangle \dS$. 
Within the Stokes limit the resulting equations have been derived in \cite{Arroyoetal_PRE_2009,Salbreuxetal_PRE_2017,Torres-Sanchezetal_JFM_2019} and are numerically solved for simply-connected and axisymmetric surfaces in \cite{Torres-Sanchezetal_JFM_2019} and \cite{Arroyoetal_PRE_2009,Mietkeetal_PNAS_2019}, respectively. We will consider the full surface Navier-Stokes equations and provide a numerical approach for general surfaces (not necessarily simply-connected). Eqs. \eqref{eq:sns1} and \eqref{eq:sns2} are not affected by the considered extensions but eq. \eqref{eq:sns3} changes for $\phi = 0$ to  
\begin{align}
	\label{eq:sns3*}
    \partial_t\vnor + 2\levicivita{\tangentVelocity}{\vnor} + \surfaceInnerProduct{\shop\tangentVelocity}{\tangentVelocity} &= -\pressure\meanc + \frac{2}{\Re}\left( \surfaceInnerProduct{\shop}{\GradSurf\tangentVelocity}  - \vnor\|\shop\|^2\right) \nonumber \\
    &\qquad + \frac{1}{\Be} \left(-\laplaceBeltrami\meanc - \frac{1}{2}\meanc^3  + 2\meanc\gaussc\right) 
    %+ \frac{\volPen}{\volZero^2} (\vol - \volZero)
\end{align}
with $\laplaceBeltrami$ Laplace-Beltrami operator. 

\section{Numerical approach}
\label{sec3}

To numerically solve eqs. \eqref{eq:sns1}, \eqref{eq:sns2} and \eqref{eq:sns3*} we consider a semi-implicit Euler timestepping scheme, a Chorin-like projection approach for eqs. \eqref{eq:sns1} and \eqref{eq:sns2}, similar to  \cite{Reutheretal_PF_2018,Nitschkeetal_PRF_2019}, evolution of geometric quantities and the generic finite element approach proposed in \cite{Nestleretal_JCP_2019}. The latter is based on a reformulation of all operators and quantities in Cartesian coordinates and penalization of normal components. Other applications of this approach can be found in, \eg, \cite{Nestleretal_JNS_2018,Jankuhnetal_IFB_2018,Olshanskiietal_SIAMJSC_2018,Nitschkeetal_PRSA_2018,Grossetal_SIAMJNA_2018,Hansboetal_IMAJNA_2020} for stationary and \cite{Nitschkeetal_arXiv_2019} for evolving surfaces.

\subsection{Timediscretization}

Let $0 = t^0 < t^1 < t^2 < \dots$ be a partition of the time with timestep width $\tau^m:=t^{m}-t^{m-1}$.
Each variable/quantity with a superscript index $m$ corresponds to the respective variable/quantity at time $t^m$. 
The overall algorithm for eqs. \eqref{eq:sns1}, \eqref{eq:sns2} and \eqref{eq:sns3*} reads as follows: 
for $m=1,2,\dots$ do 
\begin{enumerate}
    \item Move the geometry according to $\partial_t\para = \vnor\normal$, which reads in the timediscrete setting
        \begin{align}
            \label{eq:num1}
            \para^{m} &= \para^{m-1} + \tau^m\vnor^{m-1}\normal^{m-1}
        \end{align} 
        with the parametrization of the initial geometry $\para^{0}$ and corresponding initial normal vector $\normal^{0}$.
    \item Update the normal vector according to $\partial_t\normal = -\GradSurf\vnor$, see \cite{Huisken_JDG_1984}. This reads in the timediscrete setting
        \begin{align}
            \label{eq:num2}
            \normal^{m} &= \normal^{m-1} - \tau^m\GradSurf\vnor^{m-1}.
        \end{align}
    \item Update all other geometric quantities, \eg\ the mean curvature $\meanc^{m}$, the Gaussian curvature $\gaussc^{m}$, the projection $\surfaceId^{m}$ and the shape operator $\shop^{m}$, 
    %and the volume $\vol^m$
    by using the computed normal vector $\normal^{m}$. For convergence tests of this approach we refer to \cite{Nitschkeetal_arXiv_2019}.
    \item Solve for normal velocity $\vnor^m$, intermediate tangential velocity $\intermediateVelocity$ and  pressure $\pressure^m$ 
        \begin{align}
            \label{eq:num3}
            \textup{d}_{\vnor}^{m} + 2\levicivita{\tangentVelocity^{m}}{\vnor^m}  &= - \frac{2}{\Re}\vnor^m\|\shop^m\|^2 + \abbrevG^m \\
            \label{eq:num4}
            \mathbf{d}_{\tangentVelocity}^* + \levicivita{\tangentVelocity^{*}}{\intermediateVelocity} - \vnor^m\shop^m\intermediateVelocity &= \frac{1}{\Re} ( \laplaceBochner\intermediateVelocity + \gaussc^m\intermediateVelocity) + \f^m \\
            \label{eq:num5}
    			-\tau^m\laplaceBeltrami\pressure^m &= -\DivSurf\intermediateVelocity + \vnor^m\meanc^m.
        \end{align}
        with discrete time-derivatives $\textup{d}_{\vnor}^{m} = \frac{1}{\tau^m} (\vnor^m - \vnor^{m-1})$ and $\mathbf{d}_{\tangentVelocity}^* = \frac{1}{\tau^m} \surfaceId^m (\intermediateVelocity - \tangentVelocity^{m-1})$ and coupling terms $\abbrevG^m = - \pressure^{m}\meanc^m + \frac{2}{\Re}\surfaceInnerProduct{\shop^m}{\GradSurf \tangentVelocity^{m}} + \frac{1}{\Be} (-\laplaceBeltrami\meanc^m - \frac{1}{2} (\meanc^m)^3  + 2\meanc^m\gaussc^m) 
        %+ \frac{\volPen}{\volZero^2} (\vol^m - \volZero) 
        - \surfaceInnerProduct{\shop^m\tangentVelocity^{m}}{\tangentVelocity^{m}}$, 
         and $\f^m = \frac{1}{\Re}\left( \meanc^m\GradSurf\vnor^m - \vnor^m\GradSurf\meanc^m - 2\shop^m\GradSurf\vnor^m \right) + \vnor^m\GradSurf\vnor^m$. We linearize nonlinear terms in $\pressure^m$, $\vnor^m$ and $\intermediateVelocity$ around the solutions at $t^{m-1}$, \eg\ $\surfaceInnerProduct{\shop^m\tangentVelocity^{m}}{\tangentVelocity^{m}} = \surfaceInnerProduct{\shop^m\tangentVelocity^{m}}{\tangentVelocity^{m-1}} + \surfaceInnerProduct{\shop^m\tangentVelocity^{m-1}}{\tangentVelocity^{m}} - \surfaceInnerProduct{\shop^m\tangentVelocity^{m-1}}{\tangentVelocity^{m-1}}$, and the tangential velocity $\tangentVelocity^{m}$ follows from eq. \eqref{eq:num6}.
      \item Update tangential velocity $\tangentVelocity^m$ 
     		\begin{align}
                \label{eq:num6}
    			\tangentVelocity^m &= \intermediateVelocity - \tau^m\GradSurf\pressure^m.
     		\end{align}
\end{enumerate}

\subsection{Spacediscretization}

The remaining step is to discretize eqs. \eqref{eq:num3} - \eqref{eq:num5} from the above algorithm in space by using either the generic surface finite element method for tensor-valued surface PDEs proposed in \cite{Nestleretal_JCP_2019} or the surface finite element method for scalar-valued surface PDEs from \cite{Dziuketal_AN_2013}. Let $\surf_h=\surf_h(t)|_{t=t^m}$ be an interpolation of the surface $\surf=\surf(t)|_{t=t^m}$ at time $t^m$ such that $\surf_h := \bigcup_{T\in\mathcal{T}}T$, where $\mathcal{T}$ denotes a conforming triangulation. 
Furthermore, the finite element space is introduced as $\mathbb{V}(\surf_h) := \left\lbrace v\in\mathcal{C}^0(\surf_h) : v|_T\in \mathcal{P}^1(T), \forall v\in\mathcal{T} \right\rbrace$
with $\mathcal{C}^k(\surf_h)$ the space of $k$-times continuously differentiable functions on $\surf_h$ and $\mathcal{P}^l(T)$ polynomials of degree $l$ on the triangle $T\in\mathcal{T}$. 
We use the finite element space $\mathbb{V}(\surf_h)$ twice as trail and as test space and additionally introduce the $L_2$ inner product on $\surf_h$, as $(a,b) := \int_{\surf_h}\langle a, b\rangle dS$. 
Thus, the finite element approximations of eqs. \eqref{eq:num3} - \eqref{eq:num5} read: Find $\vnor^{m}\in\mathbb{V}(\surf_h)$, $\tangentVelocity^{m}\in\mathbb{V}(\surf_h)^3$ and $\pressure^{m} \in\mathbb{V}(\surf_h)$ such that $\forall \alpha \in\mathbb{V}(\surf_h)$, $\boldsymbol{\alpha} \in\mathbb{V}(\surf_h)^3$ and $\beta \in\mathbb{V}(\surf_h)$
        \begin{align}
            \label{eq:num3space}
            (\textup{d}_{\vnor}^{m},\alpha) + 2 (\levicivita{\tangentVelocity^{m}}{\vnor^m},\alpha) &= -\frac{2}{\Re}(\|\shop^m\|^2\vnor^m + \abbrevG^m + \frac{\areaPen}{\areaZero^2}(\area^{m}-\areaZero) \meanc^{m}, \alpha) \nonumber \\
            &\qquad + (D(\GradSurf \vnor^{m-1} - \GradSurf \vnor^{m}), \GradSurf \alpha) \\
            \label{eq:num4space}
            (\mathbf{d}_{\tangentVelocity}^{*}, \boldsymbol{\alpha}) + (\levicivita{\tangentVelocity^{*}}{\intermediateVelocity},\boldsymbol{\alpha}) - (\vnor^m\shop^m\intermediateVelocity,\boldsymbol{\alpha}) &= -\frac{1}{\Re} (\GradSurf\intermediateVelocity, \GradSurf\boldsymbol{\alpha}) + \frac{1}{\Re} (\gaussc^m\intermediateVelocity, \boldsymbol{\alpha}) \nonumber \\
            &\qquad + (\f^m + \omega_{\mathbf{t}} (\intermediateVelocity \cdot \normal^m) \normal^m, \boldsymbol{\alpha}) \\
		    \label{eq:num5space}
			\tau^m (\GradSurf\pressure^m, \GradSurf\beta) &= (\intermediateVelocity, \GradSurf\beta) + (\vnor^m\meanc^m, \beta).
        \end{align}
		Eq. \eqref{eq:num3space} is stabilized by artificial diffusion with coefficient $D$, following ideas of \cite{Smereka_JSC_2003} for surface diffusion. As in \cite{Nitschkeetal_arXiv_2019}, an additional term for global surface area conservation is included with a penalty parameter $\areaPen>0$, initial surface area $\areaZero$ and actual surface area $\area^m$.
		In eq. \eqref{eq:num4space} we have used the same symbols for the extended tangential velocity field  $\tangentVelocity = (\tangentVelocity_{x} \mathbf{e}^x, \tangentVelocity_{y} \mathbf{e}^y, \tangentVelocity_{z} \mathbf{e}^z)$ and the extended operators, see \cite{Reutheretal_PF_2018}. We further use $\DivSurf \tangentVelocity = \Grad \cdot \tangentVelocity - \normal \cdot (\Grad \tangentVelocity \cdot \normal)$ and introduce the additional term $(\omega_{\mathbf{t}} (\intermediateVelocity \cdot \normal^m) \normal^m,\boldsymbol{\alpha})$, with $\omega_{\mathbf{t}} > 0$, to penalize normal components of the extended tangential velocity. For convergence studies in $\omega_{\mathbf{t}}$ we refer to \cite{Nestleretal_JNS_2018}. The resulting equations for the components $\intermediateVelocity_{x}, \intermediateVelocity_{y}$ and $\intermediateVelocity_{z}$ are solved by surface finite elements.
From these fields $\tangentVelocity^m$ can be computed. For more details, especially for evaluating the local inner products in the $L_2$ inner products for the extended tangential velocity, we refer to \cite{Nestleretal_JCP_2019}. All equations are solved using the adaptive finite element toolbox AMDiS \cite{Veyetal_CVS_2007,Witkowskietal_ACM_2015}.

\begin{figure}
    \centering
    \begin{minipage}{\textwidth}
        \centering
        \def\picheight{0.19\textwidth}
        \includegraphics[height=\picheight]{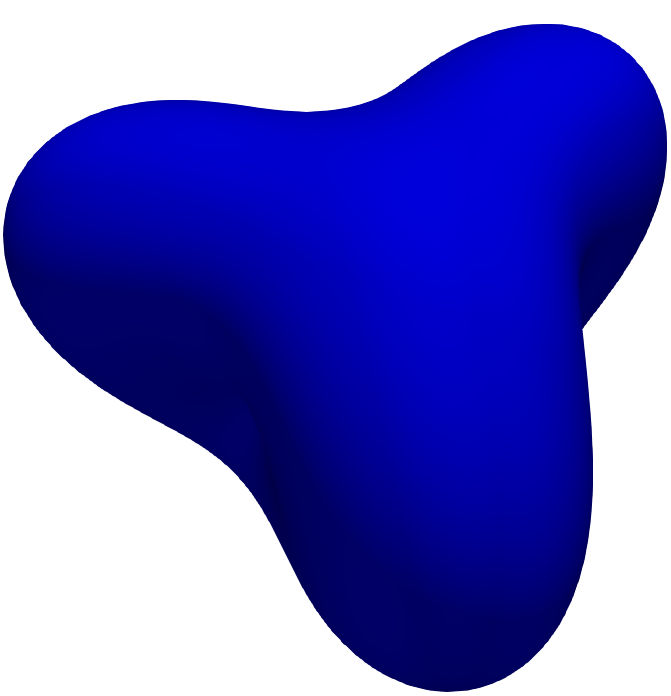}
        \includegraphics[height=\picheight]{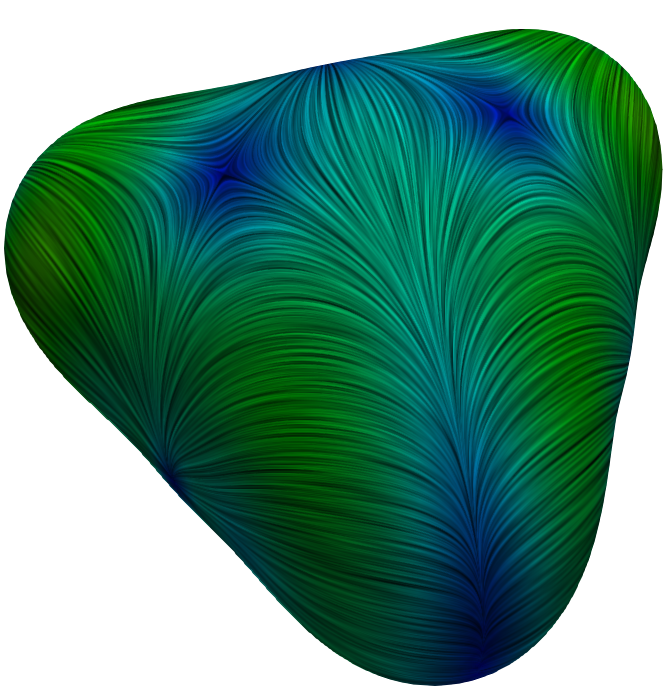}
        \includegraphics[height=\picheight]{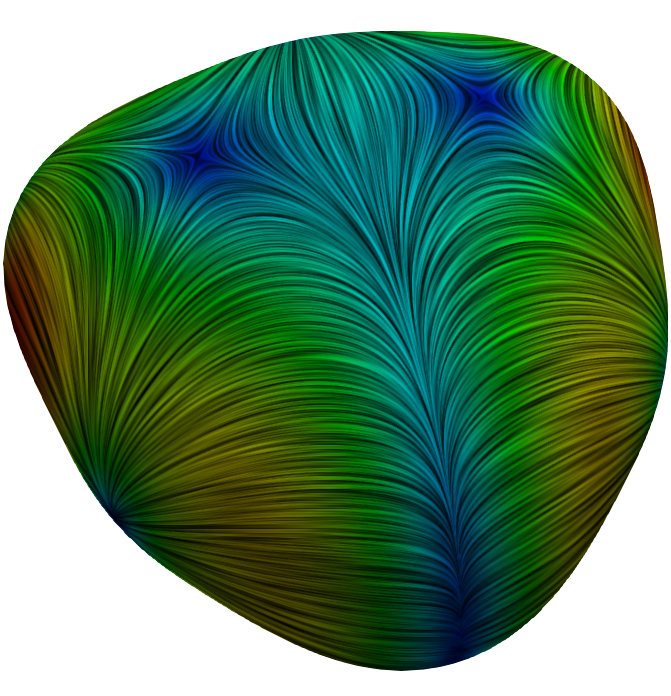}
        \includegraphics[height=\picheight]{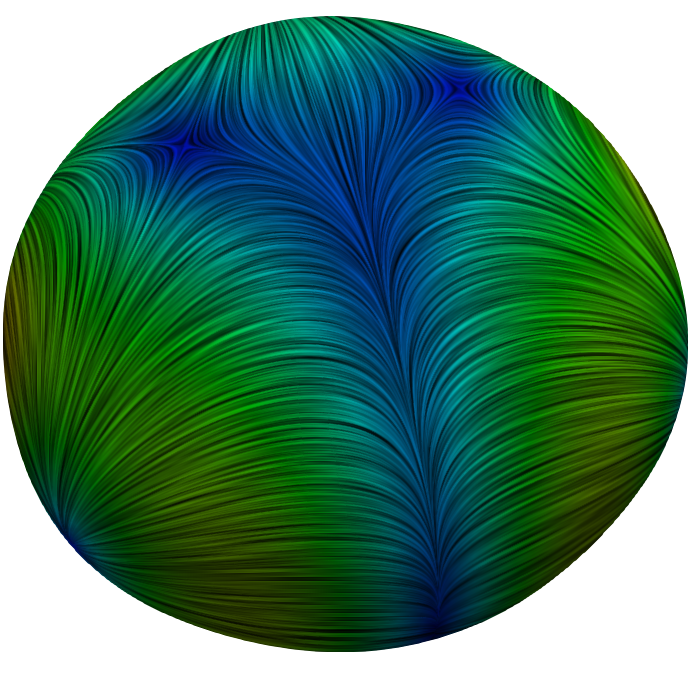}
        \includegraphics[height=\picheight]{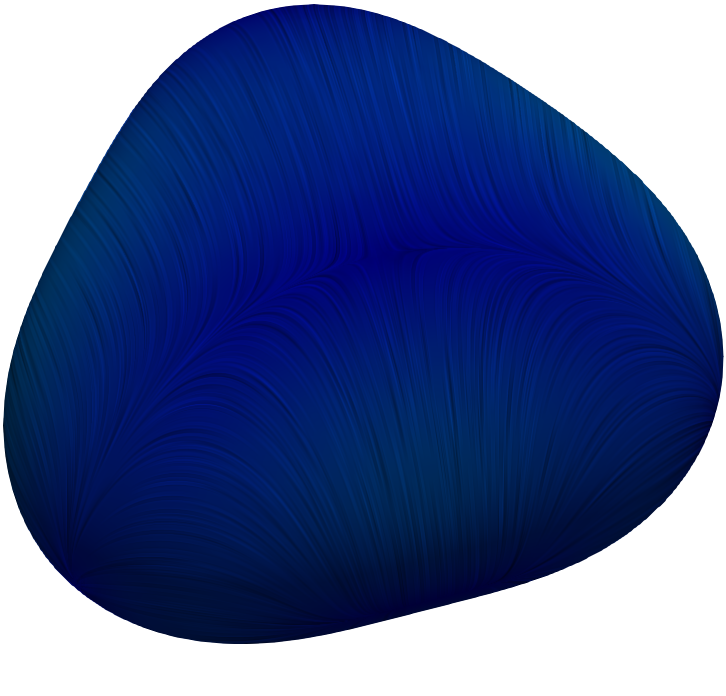}
        \includegraphics[height=\picheight]{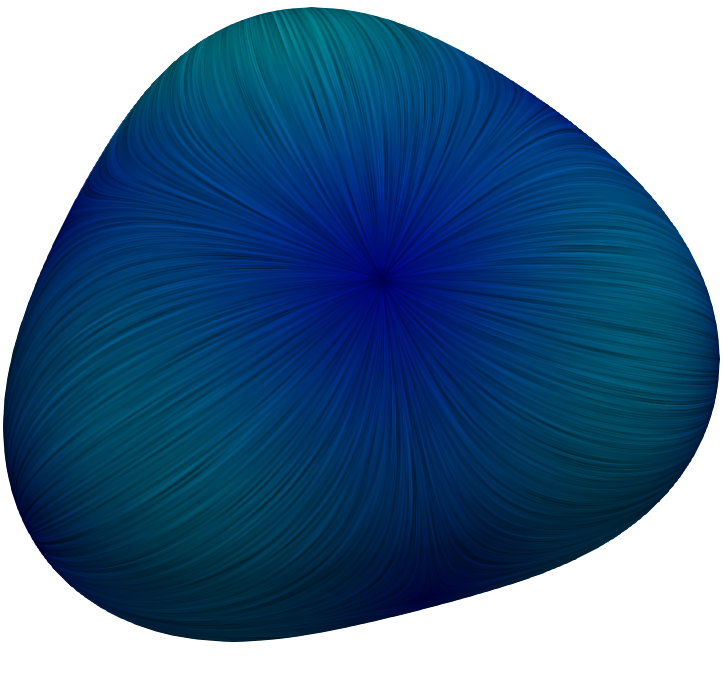}
        \includegraphics[height=\picheight]{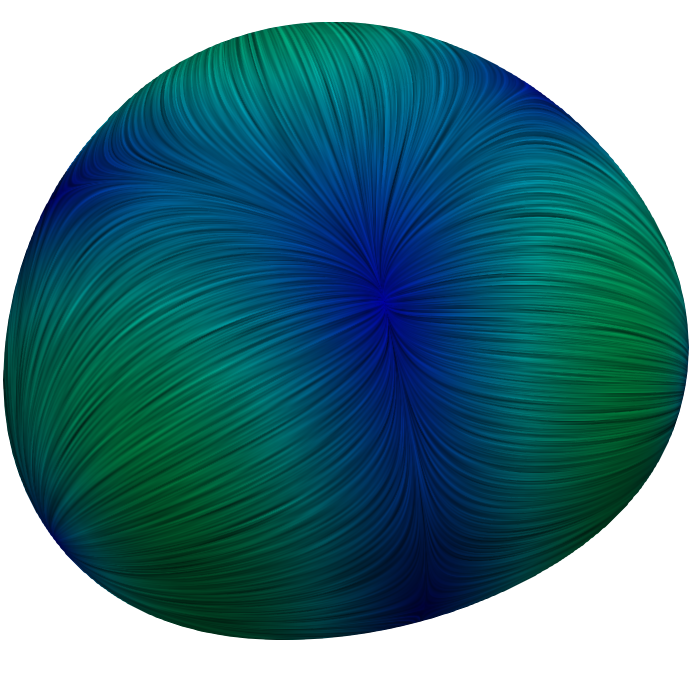}
        \includegraphics[height=\picheight]{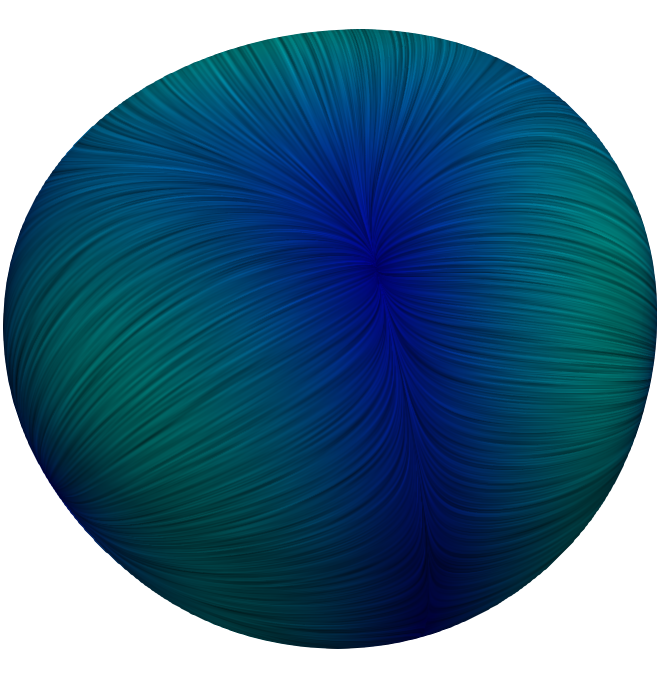}
        \includegraphics[height=\picheight]{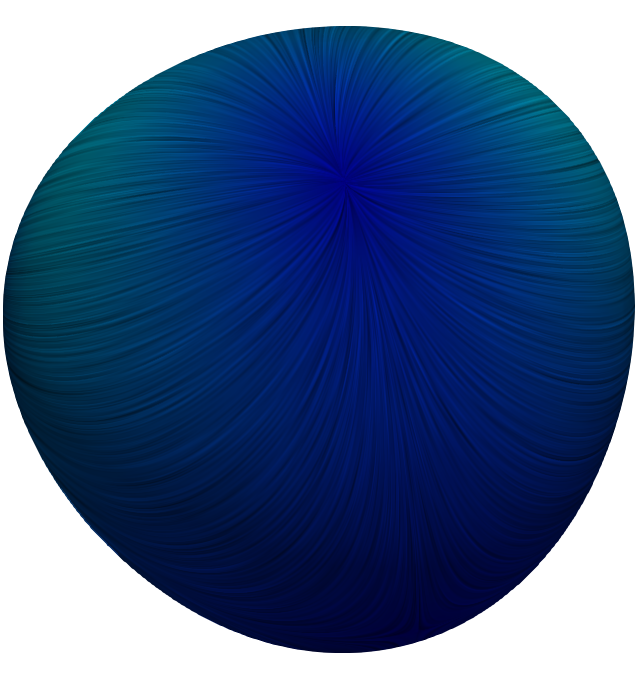}
        \includegraphics[height=\picheight]{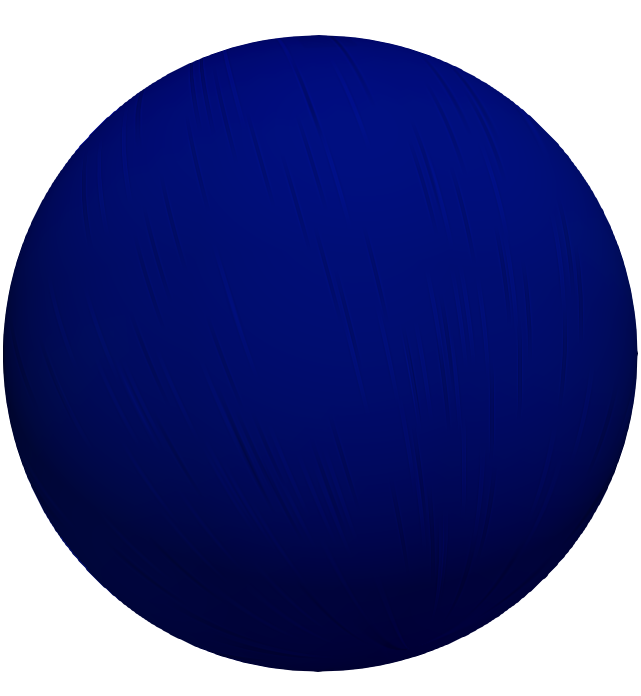}
        \insertColorbarHorizontal{$\|\tangentVelocity\|$}{$0$}{}{}{$0.4$}
    \end{minipage}
    \begin{minipage}{0.49\textwidth}
        \centering
        \input{pics/perturbedSphere/kineticEnergy.tex}
    \end{minipage}
    \begin{minipage}{0.49\textwidth}
        \centering
\definecolor{mycolor1}{rgb}{1.00000,0.00000,1.00000}%
\begin{tikzpicture}

\begin{axis}[%
width=0.7\textwidth,
height=0.4\textwidth,
at={(0,0)},
scale only axis,
xmin=0,
xmax=1.05,
xlabel style={font=\color{white!15!black}},
xlabel={$\EkinN$},
ymin=0,
ymax=0.45,
ylabel style={font=\color{white!15!black}},
ylabel={$\EkinT$},
axis background/.style={fill=white},
x label style={font=\footnotesize,at={(axis description cs:0.5,0.05)}},
y label style={font=\footnotesize,at={(axis description cs:0.05,0.5)}},
yticklabel style = {font=\footnotesize},
xticklabel style = {font=\footnotesize},
]
\addplot [color=black, line width=0.5pt, draw=none, mark size=1.5pt, mark=*, mark options={solid, fill=red}, forget plot]
  table[row sep=crcr]{%
0.0401333	0.0166829\\
0.145174	0.037538\\
0.301644	0.0437954\\
0.480155	0.0392704\\
0.655253	0.0383746\\
0.804917	0.0551565\\
0.915832	0.0972265\\
0.983074	0.16214\\
1.00856	0.238399\\
0.998157	0.310596\\
0.95911	0.364874\\
0.898244	0.392593\\
0.821276	0.391375\\
0.732775	0.364398\\
0.636537	0.318709\\
0.536125	0.263039\\
0.435328	0.205603\\
0.338309	0.152513\\
0.249323	0.10725\\
0.172148	0.0711128\\
0.109514	0.0438955\\
};
\end{axis}
\end{tikzpicture}%
    \end{minipage}
    \caption{Top/middle: Relaxation of a perturbed sphere for $t = 0$, $0.25$, $0.5$, $0.75$, $1$, $1.25$, $1.75$, $2$, $2.5$, $7.5$ (left to right, top to bottom), the tangential flow field is visualized by LIC. Bottom: Tangent, normal and overall kinetic energy $\EkinT := \frac{1}{2}\int_\surf\left\langle\tangentVelocity,\tangentVelocity\right\rangle\dS$, $\EkinN := \frac{1}{2}\int_\surf\vnor^2\dS$ and $\EkinT+\EkinN$, respectively, against time $t$ (left) and tangent kinetic energy $\EkinT$ against normal kinetic energy $\EkinN$ (right).}
    \label{fig:perturbedSphere:solution}
\end{figure}

\section{Simulation results}
\label{sec4}

All examples are chosen to demonstrate the tight coupling between $\tangentVelocity$ and $\vnor$ in the presence of curvature. The first considers a perturbed sphere with zero velocity as initial condition. The Helfrich term induces a normal velocity, which generates tangential flow. The final configuration is a sphere with zero velocity. The second examples considers a rotating Killing vector field on a sphere as initial condition. The tangential flow induces a normal velocity and with it dissipation. The final configuration is again a sphere with zero velocity. We compare the dynamics of both examples with respect to Reynolds number $\Re$ and bending capillary number $\Be$. We further consider convergence studies in meshsize $h$ and timestep width $\tau$. Besides several coarse-grained measures, such as energy components and eccentricity, eq. \eqref{eq:sns2} is used for convergence studies. It provides a severe measure for the accuracy of the algorithm, as it is never used in the approach and contains the tangential and normal parts of the velocity, $\tangentVelocity$ and $\vnor$, and the geometric quantity $\meanc$.
In the following simulations we use $D=62.5$, $\omega_{\mathbf{t}}=1\mathrm{e}{5}$ and $\areaPen=1\mathrm{e}{3}$. 

\subsection{Relaxation of perturbed sphere}
\begin{figure}
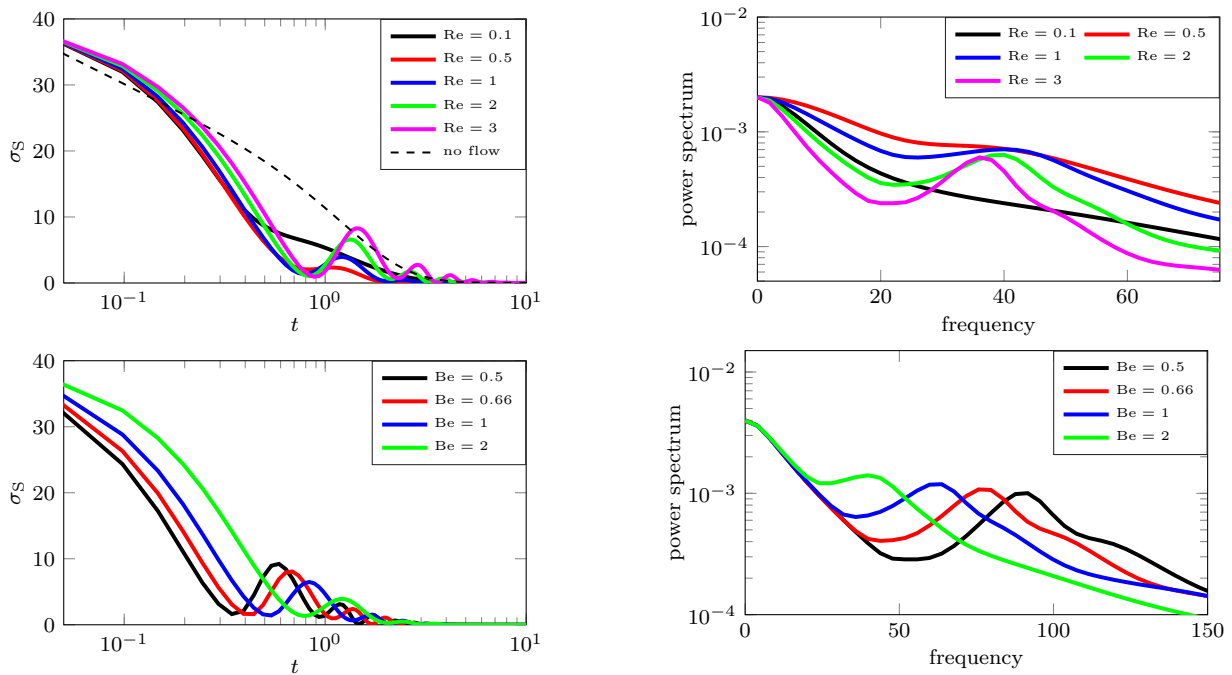

    \centering
    \begin{minipage}{0.49\textwidth}
        \centering
        \input{pics/perturbedSphere/reTest/sphericity.tex}
    \end{minipage}
        \begin{minipage}{0.49\textwidth}
        \centering
\definecolor{mycolor1}{rgb}{1.00000,0.00000,1.00000}%
\begin{tikzpicture}

\begin{axis}[%
width=0.7\textwidth,
height=0.4\textwidth,
at={(0,0)},
scale only axis,
xmin=0,
xmax=75,
xlabel style={font=\color{white!15!black}},
xlabel={$\mbox{frequency}$},
ymode=log,
ymin=5.e-5, % 1.e-4,
ymax=1.e-2,
yminorticks=true,
ylabel style={font=\color{white!15!black}},
ylabel={$\mbox{power spectrum}$},
axis background/.style={fill=white},
x label style={font=\footnotesize,at={(axis description cs:0.5,0.05)}},
y label style={font=\footnotesize,at={(axis description cs:0.05,0.5)}},
yticklabel style = {font=\footnotesize},
xticklabel style = {font=\footnotesize},
legend style={font=\tiny,at={(1.0,1.0)}, legend columns=2, anchor=north east,legend cell align=left,align=left,draw=white!15!black}, % changed
]

\addlegendentry{$\Re = 0.1$}
\addlegendentry{$\Re = 0.5$}
\addlegendentry{$\Re = 1$}
\addlegendentry{$\Re = 2$}
\addlegendentry{$\Re = 3$}

\addplot [color=black, line width=1.5pt]
  table[row sep=crcr]{%
0	0.002\\
2	0.00190628191698984\\
4	0.0016803738547141\\
6	0.00141615344300714\\
8	0.00116717953122948\\
10	0.000955294859494838\\
12	0.000786104116482874\\
14	0.000656087262029694\\
16	0.000559215407312456\\
18	0.000487434848847225\\
20	0.000433959609661719\\
22	0.000393393450038576\\
24	0.000361614796049361\\
26	0.00033624450540945\\
28	0.000315324800609963\\
30	0.000297822528602995\\
32	0.000282912235671186\\
34	0.000269908028896484\\
36	0.000258485045034349\\
38	0.000248127477409219\\
40	0.000238647097678389\\
42	0.000229779318271658\\
44	0.00022131773856073\\
46	0.000213224696168653\\
48	0.000205313527727603\\
50	0.000197618804699599\\
52	0.000190071732037125\\
54	0.000182653220820976\\
56	0.000175419309150557\\
58	0.000168304959237759\\
60	0.000161392431470969\\
62	0.000154653711378645\\
64	0.000148105472281103\\
66	0.000141791086046056\\
68	0.000135666254481217\\
70	0.000129793414346328\\
72	0.000124138073147366\\
74	0.000118714413933114\\
76	0.000113540027860742\\
78	0.000108576118301609\\
80	0.000103863615735339\\
82	9.93625567283137e-05\\
84	9.50833892505964e-05\\
86	9.10255264006362e-05\\
88	8.71558547469295e-05\\
90	8.34997647520311e-05\\
92	8.00168036372829e-05\\
94	7.67162492189815e-05\\
96	7.35877327904391e-05\\
98	7.06066267089915e-05\\
100	6.77899275922018e-05\\
102	6.51024837284399e-05\\
104	6.25559207279751e-05\\
106	6.01369577713075e-05\\
108	5.78309351083677e-05\\
110	5.565034507205e-05\\
112	5.35670708891423e-05\\
114	5.15946398767174e-05\\
116	4.97186504422224e-05\\
118	4.79320644220025e-05\\
120	4.62429175467347e-05\\
122	4.46288119529366e-05\\
124	4.31030232623879e-05\\
126	4.16501873086264e-05\\
};

\addplot [color=red, line width=1.5pt]
  table[row sep=crcr]{%
0	0.002\\
2	0.00197647246973345\\
4	0.0019097869494705\\
6	0.001809432975839\\
8	0.00168708023220553\\
10	0.00155395781483974\\
12	0.00141885985198714\\
14	0.00128818117075425\\
16	0.00116673648085433\\
18	0.00105841021374737\\
20	0.000966587443276946\\
22	0.000893612406731458\\
24	0.000840010006979603\\
26	0.000804048761115781\\
28	0.000782023127332424\\
30	0.0007692985631634\\
32	0.000761256975218692\\
34	0.000753965096612098\\
36	0.000744444296939919\\
38	0.000730666311241291\\
40	0.000711558255661271\\
42	0.000686896287089547\\
44	0.000657234785515415\\
46	0.000623747052100359\\
48	0.000587936397179352\\
50	0.000551348241783974\\
52	0.000515256828791351\\
54	0.000480538387479215\\
56	0.000447688289467977\\
58	0.00041691866410028\\
60	0.000388311494752452\\
62	0.000361883239052415\\
64	0.000337632977800508\\
66	0.000315538713260667\\
68	0.0002955322090468\\
70	0.000277505277270317\\
72	0.00026129511601896\\
74	0.000246710395723672\\
76	0.00023354752679837\\
78	0.000221604880542098\\
80	0.000210707800900944\\
82	0.00020070270432081\\
84	0.00019146873670223\\
86	0.000182909579564385\\
88	0.000174946676334835\\
90	0.000167520867352045\\
92	0.000160577745612723\\
94	0.000154073527116578\\
96	0.000147967222233579\\
98	0.000142220318133057\\
100	0.000136800772733304\\
102	0.000131675744038937\\
104	0.000126819733427225\\
106	0.000122208763189363\\
108	0.000117822418037617\\
110	0.00011364560378474\\
112	0.000109662840996087\\
114	0.000105864354759015\\
116	0.000102239849359473\\
118	9.87810881837482e-05\\
120	9.5481879935794e-05\\
122	9.23342584978233e-05\\
124	8.93335136436551e-05\\
126	8.64727679647046e-05\\
};

\addplot [color=blue, line width=1.5pt]
  table[row sep=crcr]{%
0	0.00200000000000001\\
2	0.00194617124530215\\
4	0.00180590201727073\\
6	0.00162270906616586\\
8	0.00143271634786207\\
10	0.0012559909912641\\
12	0.00110006259918421\\
14	0.000965766728186434\\
16	0.000851738221198245\\
18	0.000756747479213671\\
20	0.000681573603807161\\
22	0.000629410906164741\\
24	0.000602440024130443\\
26	0.00059677221136824\\
28	0.000604302511789164\\
30	0.000619043671309395\\
32	0.000638110372099426\\
34	0.000659435918371881\\
36	0.000680331332487195\\
38	0.000696829041498772\\
40	0.000703806317682989\\
42	0.000695601137349817\\
44	0.000668271693302077\\
46	0.00062276660868991\\
48	0.000565843727025177\\
50	0.000506980075429927\\
52	0.000453395843336206\\
54	0.000407779825602718\\
56	0.000369368999294437\\
58	0.000336115119849348\\
60	0.000306375319479805\\
62	0.000279392513550475\\
64	0.000255112158318768\\
66	0.000233744831365523\\
68	0.00021538513408205\\
70	0.000199927846600013\\
72	0.000187046084697893\\
74	0.000176306207474787\\
76	0.00016725632508549\\
78	0.000159470267810374\\
80	0.000152614576163714\\
82	0.000146423876555726\\
84	0.000140720683170881\\
86	0.000135389964641152\\
88	0.000130351623878256\\
90	0.000125565124256326\\
92	0.000120989919681179\\
94	0.000116603264259528\\
96	0.000112384659192956\\
98	0.000108314624923718\\
100	0.000104388259384976\\
102	0.000100593503045052\\
104	9.69315022122871e-05\\
106	9.34007608203726e-05\\
108	9.00000849976188e-05\\
110	8.6735221399379e-05\\
112	8.36028175481996e-05\\
114	8.06081081826099e-05\\
116	7.77497997832091e-05\\
118	7.50267726187554e-05\\
120	7.24404321265755e-05\\
122	6.99841020218391e-05\\
124	6.76578035540004e-05\\
126	6.54544861246911e-05\\
};

\addplot [color=green, line width=1.5pt]
  table[row sep=crcr]{%
0	0.002\\
2	0.00186792454695788\\
4	0.00157157590608014\\
6	0.00126356469938946\\
8	0.00100548583590778\\
10	0.000807763121992315\\
12	0.000660168274380715\\
14	0.000547324572906223\\
16	0.000462107924600273\\
18	0.00039741706958335\\
20	0.000357162803115394\\
22	0.000345267495706284\\
24	0.000348162163987176\\
26	0.000357068926790249\\
28	0.000376602631841608\\
30	0.000410162949399813\\
32	0.000457374303096506\\
34	0.000515529325270114\\
36	0.00057663461420643\\
38	0.000623669396774899\\
40	0.000629652165426317\\
42	0.00057738587977819\\
44	0.000486044934023457\\
46	0.000395852536340459\\
48	0.000329851669812429\\
50	0.000286826571285619\\
52	0.000256283477740412\\
54	0.000229521258091363\\
56	0.000203474804008482\\
58	0.000178883385955288\\
60	0.000157304899752037\\
62	0.000139544941919033\\
64	0.000125584250775779\\
66	0.000115009958319166\\
68	0.00010715965084491\\
70	0.000101404028471466\\
72	9.71068023358796e-05\\
74	9.37137356698524e-05\\
76	9.08215286383151e-05\\
78	8.81257605091414e-05\\
80	8.5510345180669e-05\\
82	8.28913060080194e-05\\
84	8.02476357018441e-05\\
86	7.75844246775168e-05\\
88	7.48879205354458e-05\\
90	7.22028811645839e-05\\
92	6.95289300021066e-05\\
94	6.68923876076424e-05\\
96	6.43070673378622e-05\\
98	6.17648522578636e-05\\
100	5.9295355660688e-05\\
102	5.68895802406693e-05\\
104	5.4573540622954e-05\\
106	5.23583871650652e-05\\
108	5.02453765143484e-05\\
110	4.82572933296694e-05\\
112	4.6378810840368e-05\\
114	4.46229124004989e-05\\
116	4.29807277040868e-05\\
118	4.14433704766433e-05\\
120	4.00161564336341e-05\\
122	3.86772499650924e-05\\
124	3.74331211830272e-05\\
126	3.62684048820635e-05\\
};

\addplot [color=mycolor1, line width=1.5pt]
  table[row sep=crcr]{%
0	0.002\\
2	0.00178984574138391\\
4	0.00136859417355835\\
6	0.00100415828132783\\
8	0.000742025213015542\\
10	0.000564524140348528\\
12	0.000443168826415546\\
14	0.00035412924748845\\
16	0.000290248393059876\\
18	0.000249751051968626\\
20	0.000238648512674815\\
22	0.000238620723815549\\
24	0.000243012968559271\\
26	0.000265549338181446\\
28	0.000306365956775247\\
30	0.000369776860188709\\
32	0.000451894477497272\\
34	0.000539676970509469\\
36	0.000598565883132733\\
38	0.000567768883824812\\
40	0.00045403423218369\\
42	0.000336703982042134\\
44	0.000263758066924584\\
46	0.000227151515208136\\
48	0.000203506865137895\\
50	0.000180077455505698\\
52	0.000155196291395781\\
54	0.000132147735206578\\
56	0.000113371673448944\\
58	9.84133514124027e-05\\
60	8.72875743854208e-05\\
62	7.92402416715388e-05\\
64	7.36704597629814e-05\\
66	7.02875334791726e-05\\
68	6.79904138336829e-05\\
70	6.65011534362324e-05\\
72	6.50832656464327e-05\\
74	6.3535895933069e-05\\
76	6.1930800846303e-05\\
78	5.99521382533267e-05\\
80	5.79888247656084e-05\\
82	5.58536604850864e-05\\
84	5.37197421368115e-05\\
86	5.16936274640318e-05\\
88	4.95532881481232e-05\\
90	4.75670834825889e-05\\
92	4.54894745562697e-05\\
94	4.34724535664821e-05\\
96	4.15538109216015e-05\\
98	3.9616639021162e-05\\
100	3.78958040494464e-05\\
102	3.61843410086936e-05\\
104	3.46325030074807e-05\\
106	3.32027725265335e-05\\
108	3.18145150552895e-05\\
110	3.06213401572063e-05\\
112	2.94339659496959e-05\\
114	2.83948769728404e-05\\
116	2.74264566324158e-05\\
118	2.64951114470003e-05\\
120	2.57002306822072e-05\\
122	2.48842195902579e-05\\
124	2.41829740736654e-05\\
126	2.34990131051894e-05\\
};

\end{axis}
\end{tikzpicture}%
    \end{minipage}
    \begin{minipage}{0.49\textwidth}
        \centering
        \input{pics/perturbedSphere/sphericityAlpha.tex}
    \end{minipage}
        \begin{minipage}{0.49\textwidth}
        \centering
        \input{pics/perturbedSphere/powerSpectrum.tex}
    \end{minipage}
    \caption{Deviation from a sphere $\sigma_{\mathrm{S}}$ against time $t$ for different Reynolds numbers $\Re$ (top) and Bending capillary numbers $\Be$ (bottom) for the perturbed sphere simulation (left) and power spectrum of the normalized deviation from a sphere $\sigma_{\mathrm{S}} / \langle\sigma_{\mathrm{S}}\rangle$ (right) with $\langle\sigma_{\mathrm{S}}\rangle$ the time average of $\sigma_{\mathrm{S}}$. Pure Helfrich flow is shown for comparison as dashed line and is indicated as "no flow".}
    \label{fig:perturbedSphere:Test}
\end{figure}

Let $\para_{S}(\phi, \theta)$ be the standard parametrization of the unit sphere with standard parametrization angles $\phi,\theta$. We use the parametrization
%\begin{align*}
    $\para(\phi, \theta) = r(\phi, \theta)\para_{S}(\phi, \theta)$
%\end{align*}
with a space-dependent radius $r(\phi, \theta) = 1 + r_0\cos(\phi)\sin(3\theta)$. Fig. \ref{fig:perturbedSphere:solution} shows the evolution for $r_0 = 0.4$ and zero initial velocity. The dynamics of the induced tangential flow field and shape changes are clearly visible. The correspondence between $\tangentVelocity$ and $\vnor$ is further highlighted in kinetic energy plots, with a strong increase in normal kinetic energy and an induced but delayed response of the tangent kinetic energy at the beginning. The later relaxation towards a sphere corresponds to a more intermediate coupling. The results correspond to $\Re = 1$ and $\Be = 2$ and the simulations are performed with $h = 4.68\mathrm{e}{-2}$ and $\tau = 4.9\mathrm{e}{-3}$. The dependency on $\Re$ ($\Be = 2$) and $\Be$ ($\Re = 1$) is considered in Fig. \ref{fig:perturbedSphere:Test}. The strongest oscillations are observed for large $\Re$ and small $\Be$. However, also for small $\Re$ the dynamics significantly differs from pure Helfrich flow, which is shown for comparison.

\subsection{Killing vector field}

\begin{figure}
    \centering
    \begin{minipage}{\textwidth}
        \centering
        \def\picwidth{0.19\textwidth}
        \includegraphics[width=\picwidth]{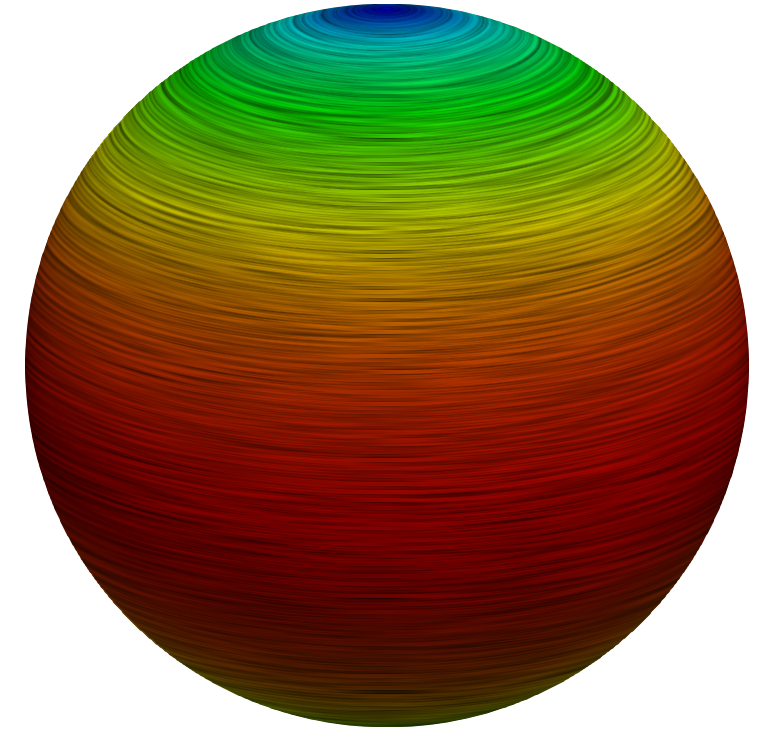}
        \includegraphics[width=\picwidth]{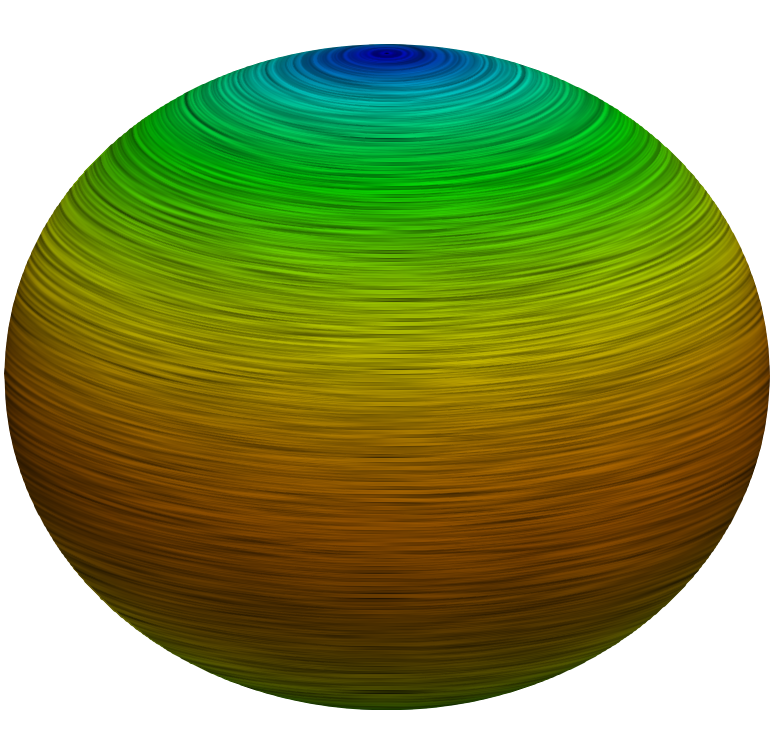}
        \includegraphics[width=\picwidth]{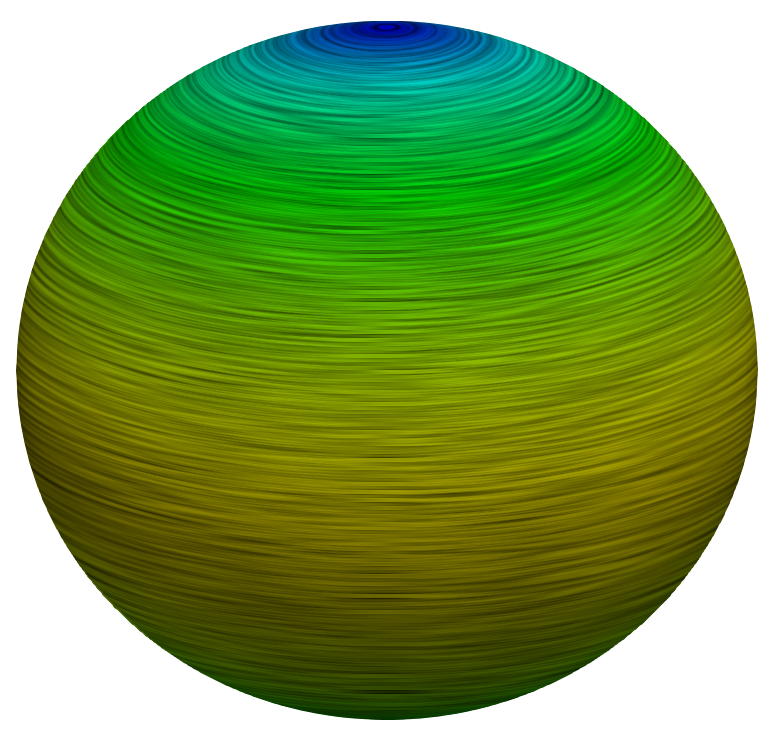}
        \includegraphics[width=\picwidth]{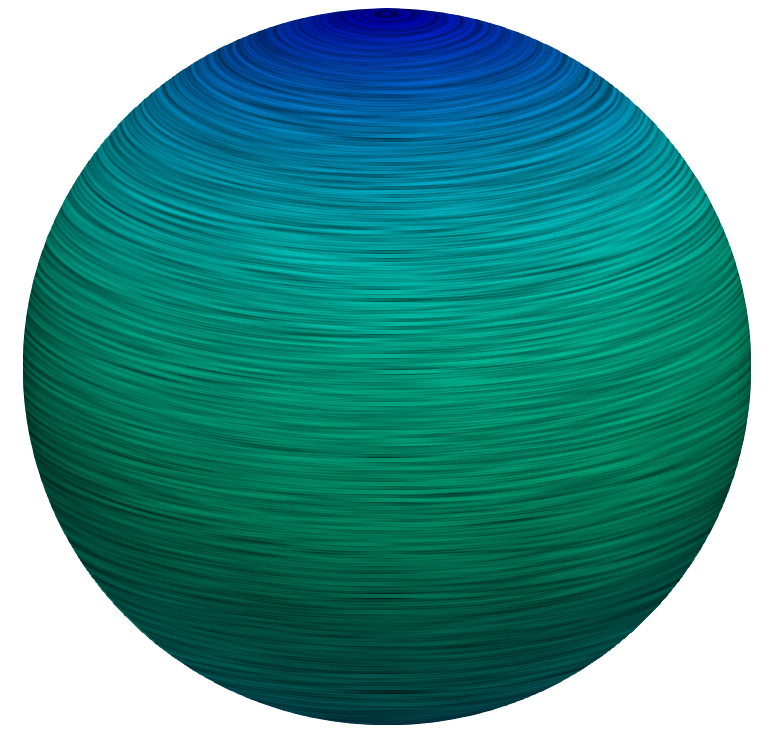}
        \includegraphics[width=\picwidth]{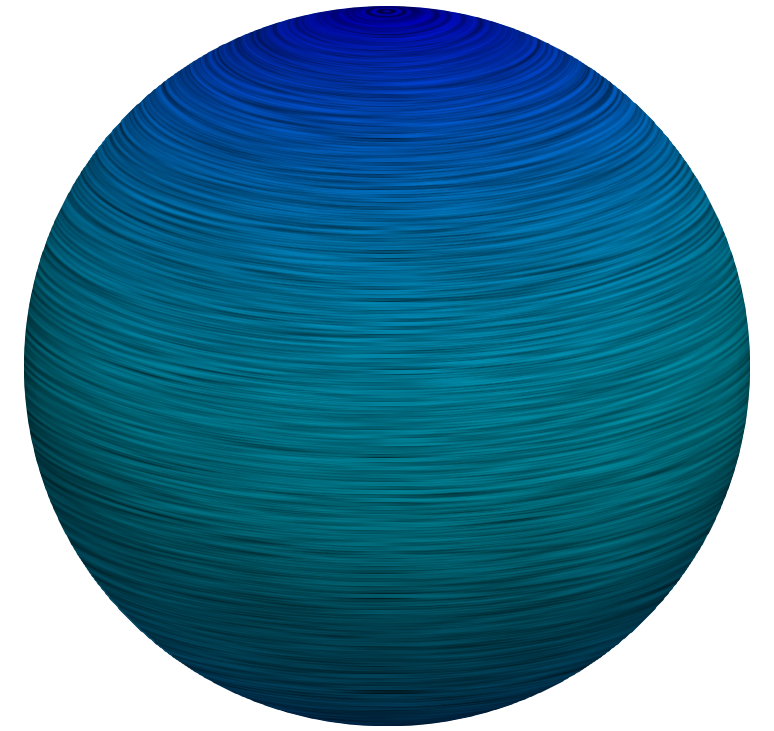}
        \insertColorbarHorizontal{$\|\tangentVelocity\|$}{$0$}{}{}{$1$}
    \end{minipage}
    \begin{minipage}{0.49\textwidth}
        \centering
        \begin{tikzpicture}
			\node (pic) at (0,0) {\includegraphics[width=0.7\textwidth]{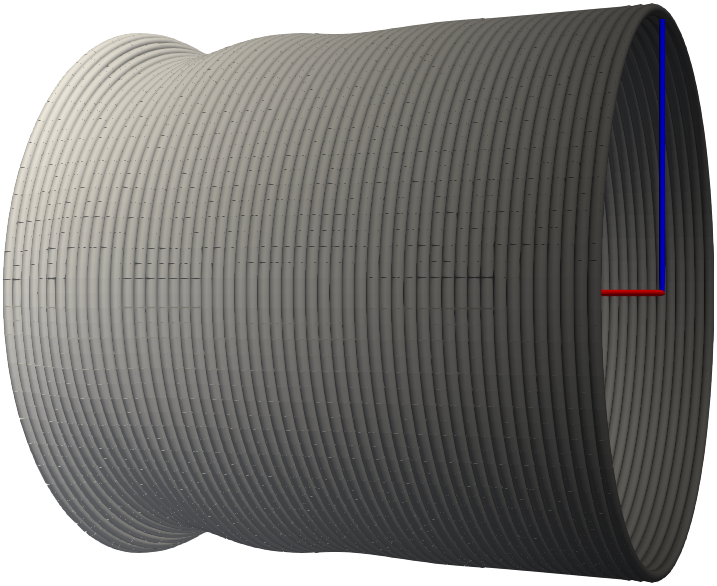}};
			\draw (1.8,0.0) node[anchor=north] {\color{white}$x$};
			\draw (2.15,1.0) node[anchor=north] {\color{white}$y$};
		\end{tikzpicture}
    \end{minipage}
    \begin{minipage}{0.49\textwidth}
        \centering
        \input{pics/killing/coordsXYPlot.tex}
    \end{minipage}
    \begin{minipage}{0.49\textwidth}
    	\centering
    	\begin{tabular}{ccccc}
    		\hline\noalign{\smallskip}
    		$h$ & $\tau$ & $e|_{t=40}$ & $\textup{EOC}_h$ & $\textup{EOC}_t$ \\
    		\noalign{\smallskip}\hline\noalign{\smallskip}
    		$7.80\mathrm{e}{-2}$ & $1.36\mathrm{e}{-2}$ & $1.72\mathrm{e}{-4}$ & $-$ & $-$ \\
    		$4.68\mathrm{e}{-2}$ & $4.90\mathrm{e}{-3}$ & $7.49\mathrm{e}{-5}$ & $1.63$ & $0.81$ \\
    		$3.34\mathrm{e}{-2}$ & $2.50\mathrm{e}{-3}$ & $4.03\mathrm{e}{-5}$ & $1.84$ & $0.92$ \\
    		$2.13\mathrm{e}{-2}$ & $1.01\mathrm{e}{-3}$ & $1.70\mathrm{e}{-5}$ & $1.91$ & $0.95$ \\
    		\noalign{\smallskip}\hline
    	\end{tabular}
    \end{minipage}
    \begin{minipage}{0.49\textwidth}
        \centering
        \input{pics/killing/sphericityH.tex}
    \end{minipage}
    \caption{(top) Relaxation of Killing vector field for $t = 0$, $1.5$, $3.5$, $15$, $20$ (left to right), the tangential flow field is visualized by LIC. (middle) Contour plot of the sliced geometry in the time interval $[0,10]$ with ascending grey scale indicating increasing time (left) and plot of the $x$-/$y$-coordinate of the geometry against time $t$ (right). (bottom) Experimental order of convergence (EOC) for different meshsizes $h$ and timestep widths $\tau$ with a constant ratio $h^2/\tau$ and the measure of the error $e := \|\DivSurf\tangentVelocity - \vnor\meanc\|_2$ (left). Deviation from a sphere $\sigma_{\mathrm{S}}$ against time $t$ with $\sigma_{\mathrm{S}} := \int_{\surf}\left(\meanc - \meanc_{\mathrm{S}}\right)^2\dS$ and $\meanc_{\mathrm{S}}$ the mean curvature of a sphere with equal surface area for different meshsizes $h$ (right).}
    \label{fig:killing:solution}
\end{figure}

\begin{figure}
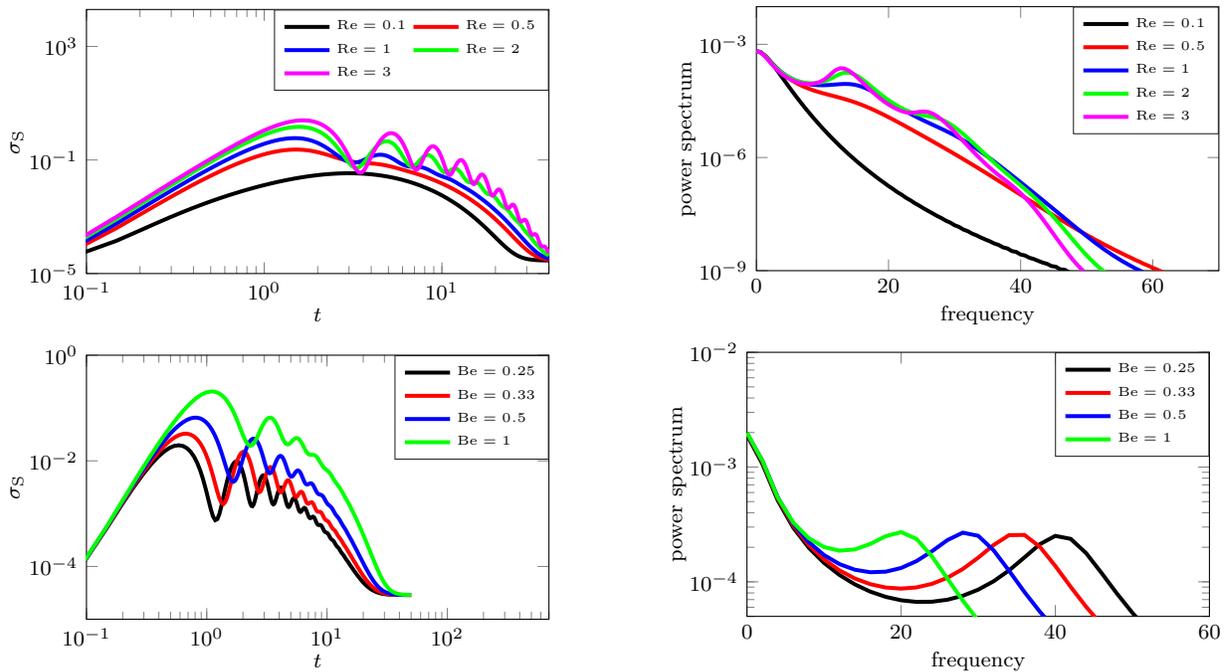

    \centering
    \begin{minipage}{0.49\textwidth}
        \centering
        \input{pics/killing/reTest/sphericity.tex}
    \end{minipage}
        \begin{minipage}{0.49\textwidth}
        \centering
        \input{pics/killing/reTest/powerSpectrum.tex}
    \end{minipage}
    \begin{minipage}{0.49\textwidth}
        \centering
        \input{pics/killing/sphericityAlpha.tex}
    \end{minipage}
        \begin{minipage}{0.49\textwidth}
        \centering
\begin{tikzpicture}

\begin{axis}[%
width=0.7\textwidth,
height=0.4\textwidth,
at={(0,0)},
scale only axis,
xmin=0,
xmax=60,
xlabel style={font=\color{white!15!black}},
xlabel={$\mbox{frequency}$},
ymode=log,
ymin=5e-05,
ymax=0.01,
yminorticks=true,
ylabel style={font=\color{white!15!black}},
ylabel={$\mbox{power spectrum}$},
axis background/.style={fill=white},
x label style={font=\footnotesize,at={(axis description cs:0.5,0.05)}},
y label style={font=\footnotesize,at={(axis description cs:0.05,0.5)}},
yticklabel style = {font=\footnotesize},
xticklabel style = {font=\footnotesize},
legend style={font=\tiny,at={(1.0,1.0)}, legend columns=1, anchor=north east,legend cell align=left,align=left,draw=white!15!black}, % changed
]

% 1
\addlegendentry{$\Be = 0.25$}
% 2
\addlegendentry{$\Be = 0.33$}
% 3
\addlegendentry{$\Be = 0.5$}
% 4
\addlegendentry{$\Be = 1$}

% 1
\addplot [color=black, line width=1.5pt]
  table[row sep=crcr]{%
0	0.0018982305002114\\
2	0.00106445889856452\\
4	0.000511353614156885\\
6	0.000296936546997919\\
8	0.00019884680687017\\
10	0.000146115287862995\\
12	0.000114741681915073\\
14	9.49506891527838e-05\\
16	8.21399973232636e-05\\
18	7.39326692970277e-05\\
20	6.90519417155487e-05\\
22	6.68421167631516e-05\\
24	6.70615638106305e-05\\
26	6.98168428484674e-05\\
28	7.55989925338668e-05\\
30	8.54293223372992e-05\\
32	0.00010114514153859\\
34	0.000125771000246103\\
36	0.000163176910065462\\
38	0.000212740474406842\\
40	0.000251341314061259\\
42	0.000236702861172169\\
44	0.000177751471707382\\
46	0.000119457735042751\\
48	7.90326399704664e-05\\
50	5.37482527401337e-05\\
52	3.8078687761673e-05\\
54	2.81418900646321e-05\\
56	2.1642882555286e-05\\
58	1.72661934910324e-05\\
60	1.42487700130834e-05\\
62	1.21369567364506e-05\\
};

% 2
\addplot [color=red, line width=1.5pt]
  table[row sep=crcr]{%
0	0.00193929114569898\\
2	0.00110754367869983\\
4	0.000531748475178015\\
6	0.000310614491146856\\
8	0.000210324412556408\\
10	0.000156992557026644\\
12	0.000125826118060229\\
14	0.000106823922882486\\
16	9.53695958108287e-05\\
18	8.91995540225759e-05\\
20	8.72798754679125e-05\\
22	8.93823806338647e-05\\
24	9.59946559049921e-05\\
26	0.000108462747645186\\
28	0.00012931168810301\\
30	0.000162283962785235\\
32	0.000209190763274197\\
34	0.000255007592136169\\
36	0.000256212377251591\\
38	0.000201920693690269\\
40	0.000137525070049297\\
42	9.03368838658776e-05\\
44	6.06105739630208e-05\\
46	4.23908458876453e-05\\
48	3.10338852490617e-05\\
50	2.37465544159052e-05\\
52	1.89378378218175e-05\\
54	1.56977102774577e-05\\
56	1.34938402866794e-05\\
58	1.20061180213776e-05\\
60	1.10343585376821e-05\\
62	1.04410304365419e-05\\
};

% 3
\addplot [color=blue, line width=1.5pt]
  table[row sep=crcr]{%
0	0.0019702013585985\\
2	0.00113551499786165\\
4	0.000544377074357134\\
6	0.000321558264982467\\
8	0.000222383382804423\\
10	0.000171079411830993\\
12	0.000142676313317165\\
14	0.000127414904079678\\
16	0.000121182689900025\\
18	0.000122573984706222\\
20	0.000131994504137378\\
22	0.000151533869605677\\
24	0.000184542709795046\\
26	0.000230761858109153\\
28	0.000268091599204686\\
30	0.000251867611985438\\
32	0.000188158597451841\\
34	0.000124907913521506\\
36	8.14978806083834e-05\\
38	5.48851549102824e-05\\
40	3.87931661339099e-05\\
42	2.88624487204061e-05\\
44	2.25637427567825e-05\\
46	1.84759476285728e-05\\
48	1.57842433975987e-05\\
50	1.39911470844425e-05\\
52	1.27417565101735e-05\\
54	1.17002021811114e-05\\
56	1.04909768773896e-05\\
58	8.82461297145631e-06\\
60	6.77757947355259e-06\\
62	4.76777099982451e-06\\
};

% 4
\addplot [color=green, line width=1.5pt]
  table[row sep=crcr]{%
0	0.00198939911307115\\
2	0.00113478644629958\\
4	0.000540672385256419\\
6	0.000328837318513158\\
8	0.000240624719941396\\
10	0.000200860030099238\\
12	0.000186757765018961\\
14	0.000191423803825338\\
16	0.00021385680229131\\
18	0.00024985256307825\\
20	0.000270951049335817\\
22	0.000236274006486652\\
24	0.000166630034582954\\
26	0.000106933736088228\\
28	6.8664835061094e-05\\
30	4.61175513692228e-05\\
32	3.28375189322667e-05\\
34	2.47488392796645e-05\\
36	1.95191636332091e-05\\
38	1.57385819530084e-05\\
40	1.24896455050854e-05\\
42	9.34324982478147e-06\\
44	6.43232249404338e-06\\
46	4.10434065598332e-06\\
48	2.49079818860694e-06\\
50	1.47687641337222e-06\\
52	8.72109838261738e-07\\
54	5.18044726951426e-07\\
56	3.10344335843993e-07\\
58	1.86996544480796e-07\\
60	1.12709494368824e-07\\
62	6.76329019739499e-08\\
};

\end{axis}
\end{tikzpicture}%
    \end{minipage}
    \caption{Deviation from a sphere $\sigma_{\mathrm{S}}$ against time $t$ for different Reynolds numbers $\Re$ (top) and Bending capillary numbers $\Be$ (bottom) for the Killing vector field relaxation (left) and power spectrum of the normalized deviation from a sphere $\sigma_{\mathrm{S}} / \langle\sigma_{\mathrm{S}}\rangle$ (right) with $\langle\sigma_{\mathrm{S}}\rangle$ the time average of $\sigma_{\mathrm{S}}$.}
    \label{fig:killing:alphaTest}
\end{figure}

The initial tangential velocity on the unit sphere is given by the Killing vector field $\tangentVelocity|_{t=0}=\left(-z,0,x\right)^T$ with $\para=\left(x,y,z\right)^T\in\surf$. The tangential velocity induces deformations towards ellipsoidal-like shapes. Due to the induced normal velocity energy dissipates. Theoretically a force balance with the bending forces of the Helfrich energy can be established. Using the axisymmetric setting an ordinary differential equation for this meta-stable states can be derived. However, these states can never be reached during evolution. The shape instead overshoots, oscillates and further dissipates energy, which decreases the driving force and leads to a relaxation back to a sphere with zero tangential velocity, see Fig. \ref{fig:killing:solution}. The results correspond to $\Re = 1$ and $\Be = 2$ and the simulations are performed with $h = 4.68\mathrm{e}{-2}$ and $\tau = 4.9\mathrm{e}{-3}$.
Convergence studies with respect to $h$ and $\tau$ are considered, indicating almost second order convergence in $h$ and first order in $\tau$. However, also number, time and strength of shape oscillations change with refinement, see Fig. \ref{fig:killing:solution}(bottom, right). The dependency of the dynamics on $\Re$ ($\Be = 2$) and $\Be$ ($\Re = 1$) is shown in Fig. \ref{fig:killing:alphaTest}, again with $h = 4.68\mathrm{e}{-2}$ and $\tau = 4.9\mathrm{e}{-3}$. The results are similar to Fig. \ref{fig:perturbedSphere:Test}.


\section{Conclusion}
\label{sec5}

With the considered thin film limit we have provided a new approach to derive the governing equations of fluid deformable surfaces. They consider fluid-like behaviour in tangential and normal direction beyond the Stokes limit and are supplemented by a Helfrich energy to model solid-like (bending) behaviour in normal direction. The splitting of the surface velocity in tangential and normal components shows the tight interplay of them with geometric quantities of the surface. This is known for the rate-of-deformation tensor. However, additional coupling terms are also present in the inertial terms. The considered numerical approach to solve these equations, which combines evolution of geometric quantities with surface finite elements and a general finite element method for tangential tensor-valued surface partial differential equations, is applicable to general surfaces (not restricted to simply-connected surfaces) and shows reasonable convergence properties with respect to meshsize $h$ (2nd order) and timestep width $\tau$ (1st order). The computational examples are chosen to demonstrate the coupling between tangential and normal velocities, where in the presence  of  curvature  any  shape  change  is  accompanied  by  a  tangential  flow  and, vice versa, the surface deforms due to tangential flow. The dynamics of the relaxation strongly depend on the fluid and solid properties. However, the simulations also show that Killing vector fields are only possible as meta-stable states, in situations where the viscous force is balanced by the bending force. The only possible stable stationary state in the considered setting is a sphere with zero velocity.

The computational examples can provide benchmark problems for other numerical approaches, which can be extended to the considered model, \eg\ \cite{Nitschkeetal_book_2017,Olshanskiietal_SIAMJSC_2018,Torres-Sanchezetal_JCP_2020,Ledereretal_arXiv_2019} and form the basis for more complex models, which include coupling with concentration fields for proteins and dependency of $\meanc_0$ on concentration in lipid bilayers, or coupling with liquid crystal theory as in \cite{Nitschkeetal_PRF_2019} for Erickson-Leslie type models or with Landau-de Gennes theory on surfaces \cite{Nitschkeetal_arXiv_2019} for Beris-Edwards type models, which also can be extended by active contributions to model, \eg, phenomena as considered in \cite{Keberetal_Science_2014}. However, any quantitative comparison in these applications will require to also consider the surrounding bulk phases, at least a constraint for the enclosed volume. Even if the approach is applicable for general surfaces it cannot handle topological changes. This would require a reformulation of the equations using an implicit description, \eg, the diffuse interface approach \cite{Raetzetal_CMS_2006}.

Acknowledgements. AV was supported by DFG through FOR3013. We further acknowledge computing resources provided by JSC withing HDR06 and ZIH at TU Dresden.

Declaration of Interests. The authors report no conflict of interest.

\bibliography{bib}
\bibliographystyle{jfm}

\end{document}